\documentclass[superscriptaddress,nofootinbib,twocolumn]{revtex4-2}
\usepackage{amssymb}
\usepackage{amsmath}
\usepackage{graphicx}
\usepackage{hyperref}
\usepackage{mathdots}
 \usepackage{xcolor}
\usepackage{slashed}
\usepackage{verbatim}
\usepackage{cancel}

\makeatletter
\def\@hangfrom@section#1#2#3{\@hangfrom{#1#2}#3}
\def\@hangfroms@section#1#2{#1#2}
\makeatother

\newcommand{\be}{\begin{equation}}
\newcommand{\ee}{\end{equation}}
\newcommand{\bea}{\begin{eqnarray}}
\newcommand{\eea}{\end{eqnarray}}

\makeatletter
\def\@ssect@ltx#1#2#3#4#5#6[#7]#8{%
  \def\H@svsec{\phantomsection}%
  \@tempskipa #5\relax
  \@ifdim{\@tempskipa>\z@}{%
    \begingroup
      \interlinepenalty \@M
      #6{%
       \@ifundefined{@hangfroms@#1}{\@hang@froms}{\csname @hangfroms@#1\endcsname}%
       {\hskip#3\relax\H@svsec}{#8}%
      }%
      \@@par
    \endgroup
    \@ifundefined{#1smark}{\@gobble}{\csname #1smark\endcsname}{#7}%
  }{%
    \def\@svsechd{%
      #6{%
       \@ifundefined{@runin@tos@#1}{\@runin@tos}{\csname @runin@tos@#1\endcsname}%
       {\hskip#3\relax\H@svsec}{#8}%
      }%
      \@ifundefined{#1smark}{\@gobble}{\csname #1smark\endcsname}{#7}%
      \addcontentsline{toc}{#1}{\protect\numberline{}#8}%
    }%
  }%
  \@xsect{#5}%
}%
\makeatother

\begin{document}

\title{A Proposal for symTFTs and symThs from Holography}

\author{Francesco Mignosa}\email{francesco.mignosa02@gmail.com}\author{Diego Rodriguez-Gomez}\email{d.rodriguez.gomez@uniovi.es}
\affiliation{Department of Physics, Universidad de Oviedo, \\ C/ Federico Garc\'ia Lorca  18, 33007  Oviedo, Spain}\affiliation{Instituto Universitario de Ciencias y Tecnolog\'ias Espaciales de Asturias (ICTEA),\\  C/~de la Independencia 13, 33004 Oviedo, Spain.}

\begin{abstract}
We propose an embedding of the symTFT construction in (finite cut-off) holography. The proposal passes several non-trivial consistency checks reproducing the expected symTFTs in various cases, including the recently discussed symTFT for the 1-form symmetry of 4d $\mathcal{N}=4$ SYM. Moreover, we comment on the possibility of unifying the symTFT and the symTh descriptions via the democratic formulation of Supergravity, using the 4d $\mathcal{N}=4$ SYM theory as an example.
\end{abstract}

\maketitle

\tableofcontents

\makeatletter
\let\toc@pre\relax
\let\toc@post\relax
\makeatother 

\section{Introduction}

Symmetries play a crucial role in Physics. In recent years, there has been intense research investigating the very notion of symmetry and its implications. This led to the developement of new tools adapted to analyze symmetries in isolation, separating them from the dynamics of the corresponding Quantum Field Theory (QFT), as well as possible variants of the theories themselves obtained by flat gaugings of global symmetries. This structure is captured by the so-called symmetry Topological Field Theory (symTFT) construction \cite{Freed:2012bs,Apruzzi:2021nmk}. This is a Topological Quantum Field Theory (TQFT) defined on a slab extended in one extra dimension that captures the symmetries of the original theory and their possible gaugings.

This construction immediately reminds of holography. It is then very tempting, for holographic theories, to look for the symTFT within the holographic dual (see \cite{Bonetti:2024cjk, Argurio:2024oym, Cvetic:2024dzu,Yu:2024jtk, Gagliano:2024off} for related work along these lines). Yet, upon closer inspection, a precise identification appears to be much less obvious. Through the standard holographic correspondence, the $p+1$-form gauge fields of the gravitational theory and their Wess-Zumino (WZ) couplings capture the symmetries of the dual QFT and their anomalies respectively. This is, in essence, the so-called symmetry Theory (symTh) of \cite{Apruzzi:2024htg} rather than a TQFT.
Moreover, the standard holographic prescription also identifies the exchange of Dirichlet and Neumann boundary conditions for the gauge fields in $AdS$ with the dynamical gauging of a global symmetry in the dual field theory. This raises two immediate issues. On one hand, upon dynamical gauging, a Conformal Field Theory (CFT) generically results in a non-conformal theory, which, in the standard holographic prescription, translates in the non-normalizability of the corresponding boundary condition. Thus, strictly speaking, generically it is not even well-defined to consider arbitrary boundary conditions within the standard holographic prescription. On top, even if one blindly allows for arbitrary boundary conditions in the symTh, the dual CFT exhibits completely different global symmetries (for example, in a 4d CFT a $U(1)^{(0)}$ symmetry gives rise to a $U(1)^{(1)}$ symmetry upon gauging) which correspond to a dramatically different symTFT in each case. It is thus not at all clear how to reconcile these observations. The aim of this note is to address this problem and propose an embedding of the symTFT within holography. 

\section{A proposal for S\lowercase{ym}T\lowercase{h}s and S\lowercase{ym}TFTs from holography}

Let us consider a $d$-dimensional CFT denoted by T exhibiting a $p$-form global symmetry $\Gamma^{(p)}$. We can then construct a family of theories $\mathbb{T}$ obtained by gauging (in general dynamically) arbitrary subgroups of the global symmetry of T. As a consequence, not all elements in $\mathbb{T}$ are CFTs. For example, T may have a global $U(1)^{(0)}$ form symmetry acting on matter fields; while upon gauging the whole $U(1)$, the resulting theory includes a sector with a copy of the Maxwell theory coupled to matter, which is actually an IR free QFT for $d>3$.

Suppose now that T has a holographic dual in terms of a gravitational theory in $AdS_{d+1}$. In particular, the global symmetry of T maps to the $p+1$-form gauge sector of the gravitational theory. A natural question is whether we can holographically describe other elements in $\mathbb{T}$. In the case of gauge fields in $AdS_{d+1}$, the standard holographic prescription identifies Dirichlet boundary conditions with the description of a global $U(1)^{(0)}$ of T, while Neumann boundary conditions correspond to gauging such $U(1)^{(0)}$ \cite{Witten:2003ya,Marolf:2006nd,DeWolfe:2020uzb} leading to some other element of $\mathbb{T}$. Inspired by this, one would naively  construct $\mathbb{T}$ by considering the appropriate boundary conditions for the gauge fields associated with the global symmetry. 

However, this cannot be fully correct: the $AdS/CFT$ correspondence is not well-defined for arbitrary boundary conditions. For instance, gauge fields in $AdS_{d+1}$ only admit normalizable Dirichlet boundary conditions if $d>3$ \cite{Marolf:2006nd}, which is intimately related to the fact that gauging $U(1)^{(0)}$ global symmetries results in an IR free theory (Maxwell theory coupled to matter) if $d>3$. 

To circumvent this problem, a natural way out is to impose a sharp cut-off in the radial direction $z\in[z_{\star},\infty)$, and impose arbitrary boundary conditions at the cut-off surface $z_{\star}$, which is always well-defined given that $z_{\star}$ is finite. Let us apply this reasoning to the case of T. By definition, being T a CFT, we are assuming boundary conditions such that we can send $z_{\star}$ to 0. As in \cite{Heemskerk:2010hk}, we can imagine splitting the holographic partition function as (we use $\phi$ below to denote the set of fields in the gravity side)

\begin{eqnarray}
\label{ZofT}
Z&=&\int \mathcal{D}\phi\,e^{-S_{\rm gravity}}=\\ \nonumber  &&\int \mathcal{D}\phi_{\star} \Big[\int \mathcal{D}\phi|_{z<z_{\star}}e^{-S[\phi|_{z<z_{\star}}]}\Big]\, \Big[\mathcal{D}\phi|_{z>z_{\star}}e^{-S[\phi|_{z>z_{\star}}}\Big]\\ \nonumber  &&=\int \mathcal{D}\phi_{\star} \Psi_{\rm UV}[\phi_{\star}] \, \Psi_{\rm IR}[\phi_{\star}]\,,
\end{eqnarray}
where $\phi_{\star}$ are the values of the $AdS$ fields at the surface $z_{\star}$. This equation has been taken as starting point to establish the holographic Renormalization Group flow in \cite{Heemskerk:2010hk,deBoer:1999tgo,Faulkner:2010jy}. An alternative point of view was advocated in \cite{McGough:2016lol} (see also \cite{Taylor:2018xcy,Hartman:2018tkw}), where $\Psi_{\rm IR}[\phi_{\star}]$ was identified with the partition function of the theory deformed by a particular multitrace deformation valid up to a cut-off $\Lambda\sim z_{\star}^{-1}$. Abusing of notation, we will still denote this QFT by T.\footnote{After all we are interested in the symmetries of T, and the deformation nor the cut-off will explicitly break such symmetries.} From this point of view, $\Psi_{UV}[\phi_{\star}]$ is what is needed in order to reproduce the path integral dual to the original CFT T.

In this set-up, the symmetries of T are captured by the $p+1$-form gauge field sector assuming the $AdS_{d+1}$ metric fixed. However, crucially, since $\Psi_{\rm IR}[\phi_{\star}]$ is dual to the cut-off $AdS_{d+1}^{\star}$, any boundary condition is allowed. Thus we can now legally reach other elements in $\mathbb{T}$ by a suitable choice of boundary conditions at $z_{\star}$. We propose to identify the effective theory for $p+1$-forms, with arbitrary boundary conditions, as the symTh of \cite{Apruzzi:2024htg}. In order to make the discussion more concrete, let us consider the case of a single $U(1)$ $p$-form symmetry.\footnote{The anomaly sector, captured by WZ couplings of the gravity theory, comes along for the ride as far as our present discussion is concerned, and thus we will momentarily neglect it. In Section \ref{N=4} we will be more precise, including it in order to appropriately describe the 1-form symmetries of $\mathcal{N}=4$ SYM.}    The symTh would then be\footnote{The normalization of the action ensures that the flux of field strength and of its Hodge dual are both normalized in units of $2\pi$.}
\begin{equation}
S^{\rm symTh}=\frac{1}{2\pi}\int_{AdS_{d+1}^{\star}}\frac{1}{2} dA_{p+1}\wedge \star dA_{p+1}\,,
\end{equation}
where $A_{p+1}$ is a $p+1$-form gauge field in the cut-off $AdS_{d+1}^{\star}$ for which both Dirichlet and Neumann boundary conditions are admissible. 

Note however that $\Psi_{\rm IR}[\phi_{\star}]$ is naturally regarded as a function of the fields evaluated at the cut-off surface in the gravity dual. Thus, from the point of view of $\Psi_{\rm IR}[\phi_{\star}]$ it is natural to impose always Dirichlet boundary conditions. Since duality exchanges the boundary conditions (see \textit{e.g.} \cite{DeWolfe:2020uzb}), this suggests that, in this context, a natural way to impose boundary conditions for a $p+1$-form field is to always impose Dirichlet boundary conditions either for the $p+1$-form field $A_{p+1}$ itself or the dual $d-p-2$ form field $\tilde{A}_{d-p-2}$ (which would correspond to the Neumann boundary conditions for the original field). 

While for $\Psi_{\rm IR}[\phi_{\star}]$ alone any choice of boundary conditions is allowed, in view of eq. \eqref{ZofT} it seems clear that to have a well-defined gravitational path integral, the contribution $\Psi_{\rm UV}[\phi_{\star}]$ must change accordingly. In fact, from the point of view of $\Psi_{\rm UV}[\phi_{\star}]$, the state $\Psi_{\rm IR}[\phi_{\star}]$ at the surface at $z_{\star}$ plays the role of a boundary state \cite{Faulkner:2010jy}. Thus, we expect that generically the geometry in the slab $z\in[0,z_{\star}]$  --denoted by $\mathcal{I}_{d+1}$-- will be modified. Taking into account that $\Psi_{\rm UV}[\phi_{\star}]$ can be also regarded as a function of the fixed fields at the cut-off surface, it is natural to write the theory in the modified geometry $\mathcal{I}_{d+1}$ as

\begin{equation}
\begin{cases} {\rm D:}\quad S_{\text{D}}=\frac{1}{2\pi}\int_{\mathcal{I}_{d+1}} \frac{1}{2} dA_{p+1}\wedge \star_{\mathcal{I}} dA_{p+1}\,,\\ {\rm N:}\quad S_{\text{N}}=\frac{1}{2\pi}\int_{\mathcal{I}_{d+1}} \frac{1}{2} d\tilde{A}_{d-p-2}\wedge  \star_{\mathcal{I}} d\tilde{A}_{d-p-2}\,,
\end{cases}
\end{equation}
where the $ \star_{\mathcal{I}}$ is meant to be taken with respect to the modified metric in the slab. We can separate the gauge field piece from the geometry by introducing an $\mathbb{R}$ valued $d-p$-form gauge field $f_{d-p}$ and use the identity (essentially a Hubbard-Stratonivich transformation)

\begin{align}
\label{HS}
&S=\frac{1}{2\pi}\int_{\mathcal{I}_{d+1}} \frac{1}{2} dA_q\wedge \star_{\mathcal{I}} dA_q=\nonumber \\
&=\frac{1}{2\pi}\int_{\mathcal{I}_{d+1}} \frac{1}{2}f_{d-q}\wedge  \star_{\mathcal{I}} f_{d-q}+i\,f_{d-q}\wedge dA_q\,.
\end{align}
From this point of view, only the $f_{d-p}\wedge  \star_{\mathcal{I}} f_{d-p}$ term in \eqref{HS} is sensitive to the slab geometry. We propose to interpret this term as the coupling to the physical degrees of freedom in the symTFT sense. In turn, the term linear in $f_{d-p}$ is topological, does not depend on the geometrical details of $\mathcal{I}_{d+1}$ --which hence can be regarded simply as an interval $I_{d+1}$-- and intrinsically captures the symmetries of the theory. Thus, we propose to drop the quadratic term and interpret the linear term as the symTFT. Hence, for the $p+1$ form case above, we are led to write the symTFTs depending on the chosen boundary conditions

\begin{equation}
\label{symTFTs}
\begin{cases} {\rm D:}\quad S^{\rm symTFT}_{\rm D}=\frac{i}{2\pi}\int_{\mathcal{I}_{d+1}} f_{d-p-1}\wedge dA_{p+1} \,,\\ {\rm N:}\quad S^{\rm symTFT}_N=\frac{i}{2\pi}\int_{\mathcal{I}_{d+1}}f_{p+2}\wedge d\tilde{A}_{d-p-2}\,.
\end{cases}
\end{equation}
We now recognize $S^{\rm symTFT}_{{\rm D}}/S^{\rm symTFT}_{{\rm N}}$ as the symTFT for $U(1)^{(p)}$/$U(1)^{(d-p-3)}$ symmetries as described in \cite{Antinucci:2024zjp}, in agreement with the fact that gauging a $U(1)^{(p)}$ symmetry gives back a $U(1)^{(d-p-3)}$ symmetry. Hence, we see that our procedure successfully maps the familiar exchange of boundary conditions in holography to the expected dramatic change of symTFT upon gauging.

\subsection{A closer look at the duality}

The procedure outlined above converted the traditional exchanging boundary conditions corresponding to gauging/ungauging a global symmetry into changing appropriately --and highly non-trivially-- the symTFT. We can offer another perspective on this process. Let us consider as the starting point the action $S_{\text{D}}$. One way to implement duality \cite{Rocek:1991ps,Witten:2003ya} is through gauging the shift symmetry $A_{p+1}\rightarrow A_{p+1}+\lambda_{p+1}$ by introducing a gauge field $B_{p+2}$ transforming as $B_{p+2}\rightarrow B_{p+2}+ d\lambda_{p+1}$ and couple it with a $d-p-2$ Lagrange multiplier $\tilde{A}_{d-p-2}$ ensuring that $B_{p+2}$ is closed. The action is then
\begin{align}
&\mathcal{S}=\frac{1}{2\pi}\int \frac{1}{2} (dA_{p+1}-B_{p+2})\wedge \star (dA_{p+1}-B_{p+2})+\nonumber \\
&+ i\,\tilde{A}_{d-p-2}\wedge d B_{p+2}\,.
\end{align}
Integrating out the field $\tilde{A}_{d-p-2}$ gives then back $S_{\rm D}$. In turn, integrating out $A_{p+1}$ leads to $S_{\text{N}}$. Moreover, this theory can be regarded as the low-energy limit\footnote{A quick way to argue for this is to note for instance that upon gauge-fixing $A_{p+1}=0$ in \eqref{Shat}, $B_{p+2}$ is simply a massive gauge field with standard kinetic term. Setting it to zero leaves us with $\widehat{\mathcal{S}}=1/2 \int d\tilde{A}_{d-p-2}\wedge \star d\tilde{A}_{d-p-2}$, which is simply the theory in the dual variables.}  of a theory equipped with kinetic terms for $\tilde{A}_{d-p-2}$ and $B_{p+2}$, namely 
\begin{eqnarray}
\label{Shat}
 \widehat{\mathcal{S}}&=&\frac{1}{2\pi}\int \frac{1}{2} (dA_{p+1}-B_{p+2})\wedge \star (dA_{p+1}-B_{p+2}) \nonumber \\ && +\frac{1}{2}dB_{p+2}\wedge \star dB_{p+2}  + \frac{1}{2} d\tilde{A}_{d-p-2}\wedge \star d\tilde{A}_{d-p-2}\nonumber \\ && +i\, \tilde{A}_{d-p-2}\wedge dB_{p+2} \,.
\end{eqnarray}
Applying our proposal to this action, we would be left with 
\begin{eqnarray}
  \widehat{\mathcal{S}}^{\rm symTFT}&=&\frac{i}{2\pi}\int h_{d-p-1}\wedge  dA_{p+1}  \\ \nonumber && - h_{d-p-1}\wedge B_{p+2} +g_{d-p-2} \wedge dB_{p+2}  \\ \nonumber &&  f_{p+2} \wedge d\tilde{A}_{d-p-2} + \tilde{A}_{d-p-2}\wedge dB_{p+2}\,.
\end{eqnarray}
We now recognize the symTFT for a $U(1)^{(p)}$ coupled to the symTFT of a gauge field (\textit{cf.} eq. (34) in \cite{Antinucci:2024zjp} for $p=0$), which explicitly reflects the gauging of the global symmetry. 

\section{Case study: 1-form symmetry in 4d $\mathcal{N}=4$ SYM}\label{N=4}

Let us now focus on the specific case of 1-form symmetries in $\mathcal{N}=4$ SYM and apply our proposed procedure. The relevant part of the Type IIB SUGRA action (reduced on the $S^5$) is

\begin{equation}
S=\frac{1}{2\pi}\int \frac{1}{2} dB_2\wedge \star dB_2 + \frac{1}{2} dC_2\wedge \star dC_2 + i\,N\, C_2\wedge dB_2\,.
\end{equation}
We can essentially import verbatim the general discussion regardless of the anomaly term.
Thus, we are led to conclude that the symTFT for $u(N)$ $\mathcal{N}=4$ SYM is

\begin{equation}
S^{\rm symTFT}_{u(N)}=\frac{i}{2\pi}\int h_2\wedge dB_2 +f_2\wedge dC_2 + N\, C_2\wedge dB_2\,,
\end{equation}
which recovers the proposal of \cite{Bergman:2025isp}. 

Let us now imagine gauging the $U(1)$ magnetic 1-form symmetry of the $u(N)$ $\mathcal{N}=4$ SYM theory. On general grounds, we expect to land on the $su(N)$ $\mathcal{N}=4$ SYM theory. Based on our previous discussion, we are led to consider the action upon dualizing $C_2$. After dualizing $C_2$ into $C_1$, we lead to

\begin{equation}
\label{Sdual}
S=\frac{1}{4\pi}\int  dB_2\wedge \star dB_2 +(dC_1+N\, B_2)\wedge \star (dC_1+N\, B_2)\,.
\end{equation}
Running our procedure now, we find

\begin{equation}
S^{\rm symTFT}_{su(N)}=\frac{i}{2\pi}\int h_2\wedge dB_2 + f_3\wedge (dC_1+N\, B_2)\,,
\end{equation}
which recovers the proposal of \cite{Bergman:2025isp} for the symTFT for $su(N)$ $\mathcal{N}=4$ SYM. 

Taking now as the starting point, eq. \eqref{Sdual}, we can imagine dualizing by gauging the $C_1\rightarrow C_1+\lambda_1$ shift symmetry by introducing a gauge field $G_2$ and a Lagrange multiplier $F_2$ imposing flatness. We end up with
\begin{eqnarray}
S&=&\frac{1}{2\pi} \int \frac{1}{2} (dC_1+N\, B_2+G_2)\wedge \star (dC_1+N\, B_2+G_2)+\nonumber  \\  &&  + \frac{1}{2} dB_2\wedge \star dB_2+i\, F_2\wedge dG_2\,.
\end{eqnarray}
Upon including kinetic terms for $G_2$ and $F_2$, using our prescription, we are led to the symTFT
\begin{align}
&S^{\rm symTFT}=\frac{i}{2\pi}\int h_2\wedge dB_2 +g_2\wedge dG_2 + f_2\wedge dF_2\nonumber \\  
&+ f_3\wedge (dC_1+N\, B_2+G_2) + F_2\wedge dG_2\,.
\end{align}

Reassuringly, this recovers the procedure in \cite{Bergman:2025isp} for going from the symTFT of $su(N)$ to the one of $u(N)$ $\mathcal{N}=4$ SYM (\textit{c.f.} eq. (4.12) in \cite{Bergman:2025isp}). 
\section{S\lowercase{ym}TFT as a limit of holography}

In \cite{Bonetti:2024cjk,Yu:2024jtk,Bergman:2025isp} it was argued that, at least in certain cases, the symTFT can arise as a limit of the symTh in standard holography --that is, in the absence of a cut-off. To begin with, note that the metric of $AdS_{d+1}$ can be written as $ds^2=z^{-2}\,ds_{I_{d+1}}^2$, with $ds_{I_{d+1}^2}=d\vec{x}^2+dz^2$ the metric of an interval. Then, for a $q$-form field $A_q$, the Hubbard-Stratonovich transformation of eq. \eqref{HS} for a $p+1$ form and its $d-p-2$ dual can be recasted as 

\begin{equation}
\begin{cases}
S_{\text{D}}=\int \frac{z^{d-2p-3}}{4\pi}f_{d-p-1}\wedge  \star_{I} f_{d-p-1}+\frac{i}{2\pi}f_{d-p-1}\wedge dA_{p+1}\,;\\
S_{\text{N}}=\int \frac{z^{3+2p-d}}{4\pi}f_{p+2}\wedge  \star_{I} f_{p+2}+\frac{i}{2\pi}f_{p+2}\wedge dA_{d-p-2}\,.
\end{cases}
\end{equation}
As a consequence, we see that depending on $d$ and $p$, as $z\rightarrow 0$ the symTh coincides with the symTFT as \cite{Bonetti:2024cjk,Yu:2024jtk,Bergman:2025isp}

\begin{enumerate}
\item $d-2p-3>0$: $S_{\text{D}}^{\rm symTh}\rightarrow S_{\text{D}}^{\rm symTFT}$
\item $d-2p-3<0$: $S_{\text{N}}^{\rm symTh}\rightarrow S_{\text{N}}^{\rm symTFT}$
\end{enumerate}
Note that cases 1 and 2 above coincide with the appropriate boundary conditions for normalizability of the $p+1$ form field in $AdS_{d+1}$. 
 Thus, with the boundary conditions selected by holography --those leading to a normalizable fluctuation in $AdS_{d+1}$-- the symTh is captured by the symTFT close to the $AdS$ boundary. This is to be expected, as in that case the cut-off of the dual QFT can be taken to infinity, resulting in a CFT which does not need a UV completion, and thus the slab geometry $\mathcal{I}_{d+1}$ above is actually simply $AdS$ again. We then see that our construction naturally accommodates the point of view in \cite{Bonetti:2024cjk,Yu:2024jtk,Bergman:2025isp}.

The case $d-2p-3=0$ is special. It corresponds to the case where both standard and alternate quantizations are allowed. Thus, symmetry alone is not enough to select boundary conditions. This is reflected in the fact that the $z$-dependence above drops in both $S_{\text{N}}$ and $S_{\text{D}}$.

To illustrate these points, let us consider $p=0$. As it is well-known \cite{Marolf:2006nd}, gauge fields in $AdS_{d+1}$ require D boundary conditions for $d>3$ and are dual to conserved currents of $U(1)^{(0)}$ symmetries in the boundary. Consistently, for D boundary conditions, $S_{\text{D}}^{\rm symTh}\rightarrow S_{\text{D}}^{\rm symTFT}$, and we recover the symTFT for a $U(1)^{(0)}$ symmetry. In turn, if $d<3$, gauge fields in $AdS_{d+1}$ require N boundary conditions and correspond to gauge fields in the dual CFT. Consistently, in this case $S_{\text{N}}^{\rm symTh}\rightarrow S_{\text{N}}^{\rm symTFT}$. The case $d=3$ is special, as it allows for both boundary conditions. Hence, symmetry alone is not enough to choose one or the other, which is reflected in the fact that the $z$ dependence drops out from the action of the symTh upon performing the Hubbard-Stratonovich transformation.

\section{Anomaly TFT from democratic formulation}
In the above Sections, we saw how holography encodes both the symTFT for a $p$-form symmetry and  for the $d-p-3$-form symmetry obtained by gauging the $p$-form symmetry itself. In particular, the two symTFT were basically related by duality in $AdS$. It is worth noticing, however, that both symTFTs can also be obtained from a single $d+2$-dimensional theory, known as the anomaly Topological Field Theory (anomalyTFT). For a $p$-form symmetry, in the absence of anomalies, the anomalyTFT is 
\begin{equation}
S^{\text{anomalyTFT}}= \frac{i}{2\pi}\int_{M_{d+2}} x_{d-p-1}\wedge d y_{p+2}\,.
\end{equation}
where we are considering the symTFT as living on a space without boundary $M_{d+1}=\partial M_{d+2}$. In order to preserve the gauge transformations $\delta x_{d-p-1}= d\lambda_{d-p-2}$ and $\delta y_{p+2}=d\lambda_{p+1}$, the bulk action must be supplemented with a boundary action. Two natural choices are
\begin{align}
& S^{\rm symTFT}_{\text{D}, \,\partial}=\frac{i}{2\pi}\, \int_{\partial M_{d+2}}  f_{d-p-1} \wedge (dA_{p+1}-  y_{p+2})\,,\nonumber \\
& S^{\rm symTFT}_{\text{N}, \,\partial}=\frac{i}{2\pi}\, \int_{\partial M_{d+2}}  \Big[(-1)^{d-p}x_{d-p-1}\wedge y_{p+2}+ \nonumber\\
&+(-1)^{d(p+1)}f_{p+2} \wedge (d\tilde{A}_{d-p-2}-  x_{d-p-1})\Big]\,,
\end{align}
leading to the two different symTFTs. Moreover, we can also supplement the anomalyTFT with two different boundary actions
\begin{align}
& S^{\rm symTh}_{\text{D}, \, \partial}= \frac{1}{4\pi}\int_{\partial M_{d+2}} y_{p+2}\wedge \star y_{p+2}\,, \nonumber\\
& S^{\rm symTh}_{\text{N}, \,\partial}=\frac{1}{4\pi}\int_{\partial M_{d+2}}\Big[2i(-1)^{d-p} x_{d-p-1}\wedge y_{p+2}+\nonumber\\ 
&+x_{d-p-1}\wedge \star x_{d-p-1}\Big]\,,
\end{align}
leading to the two symTh related by dualization in $M_{d+1}$ \cite{Maldacena:2001ss, Evnin:2023ypu, Heckman:2024oot, Heckman:2025lmw}.

It is natural to ask about the avatar of the anomalyTFT in String Theory. Our proposal, as also suggested in \cite{Gagliano:2024off}, is that the democratic formulation of supegravity realizes the anomalyTFT. This formulation treats all Supergravity fields, namely the fields and their Hodge duals, on equal footing. An immediate consequence is that duality in Supergravity is just built into this formulation. In particular the democratic formulation in 11d reads
\begin{align}
&S^{\text{demoIIA}}=\frac{i}{2\pi} \int_{M_{11}} \Big[ w_0 \, dw_{10}-w_2\wedge dw_8+w_4\wedge dw_6+\nonumber \\
&+r_3 \wedge dr_7-r_3\wedge \left(w_0\,w_8-w_2\wedge w_6+\frac{1}{2}w_4\wedge w_4\right)\Big]\,;\\
&S^{\text{demoIIB}}=\frac{i}{2\pi} \int_{M_{11}}\Big[  w_1 \wedge dw_{9}-w_3\wedge dw_7+\frac{1}{2}w_5 \wedge dw_5+\nonumber \\
&+ r_3 \wedge dr_7-r_3\wedge \left(w_1\wedge w_7-w_3\wedge w_5\right)\Big]\,.
\end{align}
where $w_p$ are the 11d avatar of the 10d field strenghts of the RR fields, while $r_3, \, r_7$ are the NSNS ones. These are $\mathbb{R}$-fields with their own gauge transfrmation. Upon reduction over the internal space, one finds the anomaly theory. Then, just as above, in the presence of a space with a boundary, the symTFT/symTh will be obtained as the theory of edge modes which cancels the bulk gauge variation of the democratic theory.  

Let us study consider in detail the case of 4d $\mathcal{N}=4$ SYM. The reduction of the democratic formulation reads
\begin{equation}
S^{\text{demo}}= \frac{i}{2\pi}\int_{M_6}\left[r_3\wedge dr_2-w_3\wedge dw_2+N r_3\wedge w_3\right]\,,
\end{equation}
with gauge symmetries $\delta r_3= d\mu_2$, $\delta w_3= d\lambda_2$, $\delta r_2= d\mu_1-N\lambda_2$ and  $\delta w_2= d\lambda_1-N\mu_2$. To cancel those, we can choose as boundary action
\begin{align}
&S^{\rm symTFT}_{\text{D},\,\partial}\hspace{-0.18cm}= \frac{i}{2\pi}\int_{\partial M_6} \hspace{-0.1cm}\Big[r_3 \wedge r_2- w_3  \wedge w_2+h_2  \wedge dB_2+f_2  \wedge dC_2\nonumber\\
&+NC_2 \wedge dB_2 -(f_2-NB_2)  \wedge w_3+ (h_2+NC_2)  \wedge r_3\Big], \nonumber \\
&S^{\rm symTh}_{\text{D},\,\partial}\hspace{-0.1cm}= \frac{1}{2\pi}\int_{\partial M_6} \Big[ir_3  \wedge r_2- iw_3  \wedge w_2+\frac{1}{2}r_3 \wedge \star r_3\nonumber \\
&+\frac{1}{2}w_3 \wedge\star w_3\Big]\,;
\end{align}
leading respectively to the $S_\text{D}^{\text{symTFT}}$ and $S_\text{D}^{\text{symTh}}$. The other variants obtained by changing boundary conditions from Dirichlet to Neumann can then be obtained by changing the edghe mode theory.
\section{Conclusions}

In this note, we have proposed an embedding of the symTFT paradigm within holography, clarifying at the same time the role of the symTh and the relation between the two in the context of the democratic formulation. Key aspects of our proposal are that a) the slab geometry of the symTFT naturally appears, b) the exchange of boundary conditions (corresponding to dynamical gauging in standard holography) of the symTh gets mapped to the change of symTFT required for consistency, and c) both the symTFT and the symTh can be obtained from the democratic formulation of Supergravity choosing the appropriate boundary conditions. 

An immediate application of our proposal is the possibility of directly constructing symTFTs for holographic theories: the symTFT can be obtained from the corresponding Supergravity lagrangian upon changing the kinetic terms for $f\wedge F$ terms or directly from the democratic formulation by choosing the appropriated boundary conditions. Along the same lines, it seems that the extension to non-abelian symmetries discussed in \cite{Bonetti:2024cjk,Apruzzi:2025hvs, Bonetti:2025dvm} is straightforward. For instance, it is natural that the symTFT for $\mathcal{N}=4$ SYM including the non-abelian $SO(6)_R$ symmetry can be directly read-off from IIB Supergravity using the consistent truncation in \cite{Baguet:2015sma}. 

It would be relevant to improve our understanding of the underlying mechanisms of our proposal, specially the role of the physical boundary which in our construction is captured by dropping the quadratic term in the Hubbard-Stratonovich transformation of eq. \eqref{HS}. It is worth noting the very intriguing fact that a similar Hubbard-Stratonovich transformation plays a relevant role in cut-off holography \cite{McGough:2016lol}.\footnote{Note that one could also write eq. \eqref{HS} as a Stuckelberg theory  $S=\int_{\mathcal{I}_{d+1}} -\frac{1}{2}(dc_{d-q-1}-f_{d-q})\wedge  \star_{\mathcal{I}} (dc_{d-q-1}-f_{d-q})+f_{d-q}\wedge dA_q$. Using the gauged shift symmetry we can fix $c_{d-q-1}=0$, recovering \eqref{HS}. The dropped term is then a kinetic term, perhaps along the lines of \cite{Witten:1998wy}.} Moreover, it would be very interesting to study the relation of the symmetry operators in holography as introduced in \cite{Bergman:2024aly,Calvo:2025kjh,Calvo:2025usj} with their symTFT counterparts under the light of the procedure in this note. We hope to come back to some of these issues in the future.

\section*{Acknowledgements}

We would like to thank O. Bergman and E. Garc\'ia-Valdecasas for many enlightening discussions and collaborations.  The authors are supported in part by the Spanish national grant MCIU-22-PID2021-123021NB-I00.

\let\oldaddcontentsline\addcontentsline
\renewcommand{\addcontentsline}[3]{}


\begin{thebibliography}{99}
\bibitem{Freed:2012bs}
D.~S.~Freed and C.~Teleman,
``Relative quantum field theory,''
\href{https://link.springer.com/article/10.1007/s00220-013-1880-1}{Commun. Math. Phys. \textbf{326} (2014), 459-476}
\href{https://arxiv.org/abs/1212.1692}{[arXiv:1212.1692 [hep-th]].}
\bibitem{Apruzzi:2021nmk}
F.~Apruzzi, F.~Bonetti, I.~Garc{\'\i}a Etxebarria, S.~S.~Hosseini and S.~Schafer-Nameki,
``Symmetry TFTs from String Theory,''
\href{https://link.springer.com/article/10.1007/s00220-023-04737-2}{Commun. Math. Phys. \textbf{402} (2023) no.1, 895-949}
\href{https://arxiv.org/abs/2112.02092}{[arXiv:2112.02092 [hep-th]].}
\bibitem{Bonetti:2024cjk}
F.~Bonetti, M.~Del Zotto and R.~Minasian,
``SymTFTs for Continuous non-Abelian Symmetries,''
\href{https://arxiv.org/abs/2402.12347}{[arXiv:2402.12347 [hep-th]].}
\bibitem{Argurio:2024oym}
R.~Argurio, F.~Benini, M.~Bertolini, G.~Galati and P.~Niro,
``On the symmetry TFT of Yang-Mills-Chern-Simons theory,''
\href{https://link.springer.com/article/10.1007/JHEP07(2024)130}{JHEP \textbf{07} (2024), 130,}
\href{https://arxiv.org/abs/2404.06601}{[arXiv:2404.06601 [hep-th]].}
\bibitem{Cvetic:2024dzu}
M.~Cveti{\v{c}}, R.~Donagi, J.~J.~Heckman, M.~H{\"u}bner and E.~Torres,
``Cornering relative symmetry theories,''
\href{https://journals.aps.org/prd/abstract/10.1103/PhysRevD.111.085026}{Phys. Rev. D \textbf{111} (2025) no.8, 085026}
\href{https://arxiv.org/abs/2408.12600}{[arXiv:2408.12600 [hep-th]].}


\bibitem{Yu:2024jtk}
X.~Yu,
``Gauging in parameter space: A top-down perspective,''
\href{https://journals.aps.org/prd/abstract/10.1103/638n-qwnm}{Phys. Rev. D \textbf{112} (2025) no.2, 025020}
\href{https://arxiv.org/abs/2411.14997}{[arXiv:2411.14997 [hep-th]].}
\bibitem{Gagliano:2024off}
F.~Gagliano and I.~Garc{\'\i}a Etxebarria,
``SymTFTs for $U(1)$ symmetries from descent,''
\href{https://arxiv.org/abs/2411.15126}{[arXiv:2411.15126 [hep-th]].}
\bibitem{Apruzzi:2024htg}
F.~Apruzzi, F.~Bedogna and N.~Dondi,
``SymTh for non-finite symmetries,''
\href{https://arxiv.org/abs/2402.14813}{[arXiv:2402.14813 [hep-th]].}


\bibitem{Witten:2003ya}
E.~Witten,
``SL(2,Z) action on three-dimensional conformal field theories with Abelian symmetry,''
\href{https://arxiv.org/abs/hep-th/0307041}{[arXiv:hep-th/0307041 [hep-th]].}

\bibitem{Marolf:2006nd}
D.~Marolf and S.~F.~Ross,
``Boundary Conditions and New Dualities: Vector Fields in AdS/CFT,''
\href{https://iopscience.iop.org/article/10.1088/1126-6708/2006/11/085}{JHEP \textbf{11} (2006), 085}
\href{https://arxiv.org/abs/hep-th/0606113}{[arXiv:hep-th/0606113 [hep-th]].}

\bibitem{DeWolfe:2020uzb}
O.~DeWolfe and K.~Higginbotham,
``Generalized symmetries and 2-groups via electromagnetic duality in $AdS/CFT$,''
\href{https://journals.aps.org/prd/abstract/10.1103/PhysRevD.103.026011}{Phys. Rev. D \textbf{103} (2021) no.2, 026011}
\href{https://arxiv.org/abs/2010.06594}{[arXiv:2010.06594 [hep-th]].}


\bibitem{Heemskerk:2010hk}
I.~Heemskerk and J.~Polchinski,
``Holographic and Wilsonian Renormalization Groups,''
\href{https://link.springer.com/article/10.1007/JHEP06(2011)031}{JHEP \textbf{06} (2011), 031}
\href{https://arxiv.org/abs/1010.1264}{[arXiv:1010.1264 [hep-th]].}
\bibitem{deBoer:1999tgo}
J.~de Boer, E.~P.~Verlinde and H.~L.~Verlinde,
``On the holographic renormalization group,''
\href{https://iopscience.iop.org/article/10.1088/1126-6708/2000/08/003}{JHEP \textbf{08} (2000), 003}
\href{https://arxiv.org/abs/hep-th/9912012}{[arXiv:hep-th/9912012 [hep-th]].}

\bibitem{Faulkner:2010jy}
T.~Faulkner, H.~Liu and M.~Rangamani,
``Integrating out geometry: Holographic Wilsonian RG and the membrane paradigm,''
\href{https://link.springer.com/article/10.1007/JHEP08(2011)051}{JHEP \textbf{08} (2011), 051}
\href{https://arxiv.org/abs/1010.4036}{[arXiv:1010.4036 [hep-th]].}
\bibitem{McGough:2016lol}
L.~McGough, M.~Mezei and H.~Verlinde,
``Moving the CFT into the bulk with $ T\overline{T} $,''
\href{https://link.springer.com/article/10.1007/JHEP04(2018)010}{JHEP \textbf{04} (2018), 010}
\href{https://arxiv.org/abs/1611.03470}{[arXiv:1611.03470 [hep-th]].}

\bibitem{Taylor:2018xcy}
M.~Taylor,
``$T \bar{T}$ deformations in general dimensions,''
\href{https://link.intlpress.com/JDetail/1805559900216434689}{Adv. Theor. Math. Phys. \textbf{27} (2023) no.1, 37-63}
\href{https://arxiv.org/abs/1805.10287}{[arXiv:1805.10287 [hep-th]].}

\bibitem{Hartman:2018tkw}
T.~Hartman, J.~Kruthoff, E.~Shaghoulian and A.~Tajdini,
``Holography at finite cutoff with a $T^2$ deformation,''
\href{https://link.springer.com/article/10.1007/JHEP03(2019)004}{JHEP \textbf{03} (2019), 004}
\href{https://arxiv.org/abs/1807.11401}{[arXiv:1807.11401 [hep-th]].}


\bibitem{Antinucci:2024zjp}
A.~Antinucci and F.~Benini,
``Anomalies and gauging of U(1) symmetries,''
\href{https://journals.aps.org/prb/abstract/10.1103/PhysRevB.111.024110}{Phys. Rev. B \textbf{111} (2025) no.2, 2}
\href{https://arxiv.org/abs/2401.10165}{[arXiv:2401.10165 [hep-th]].}
\bibitem{Rocek:1991ps}
M.~Rocek and E.~P.~Verlinde,
``Duality, quotients, and currents,''
\href{https://www.sciencedirect.com/science/article/abs/pii/055032139290269H?via\%3Dihub}{Nucl. Phys. B \textbf{373} (1992), 630-646}
\href{https://arxiv.org/abs/hep-th/9110053}{[arXiv:hep-th/9110053 [hep-th]].}



\bibitem{Bergman:2025isp}
O.~Bergman, E.~Garcia-Valdecasas, F.~Mignosa and D.~Rodriguez-Gomez,
``The SymTFT of $u(N)$ Yang-Mills Theory and Holography,''
\href{https://arxiv.org/abs/2508.00992}{[arXiv:2508.00992 [hep-th]].}
\bibitem{Maldacena:2001ss}
J.~M.~Maldacena, G.~W.~Moore and N.~Seiberg,
``D-brane charges in five-brane backgrounds,''
\href{https://iopscience.iop.org/article/10.1088/1126-6708/2001/10/005}{JHEP \textbf{10} (2001), 005,}
\href{https://arxiv.org/abs/hep-th/0108152}{[arXiv:hep-th/0108152 [hep-th]].}
\bibitem{Evnin:2023ypu}
O.~Evnin, E.~Joung and K.~Mkrtchyan,
``Democratic Lagrangians from topological bulk,''
\href{https://journals.aps.org/prd/abstract/10.1103/PhysRevD.109.066003}{Phys. Rev. D \textbf{109} (2024) no.6, 066003}
\href{https://arxiv.org/abs/2309.04625}{[arXiv:2309.04625 [hep-th]].}
\bibitem{Heckman:2024oot}
J.~J.~Heckman, M.~H{\"u}bner and C.~Murdia,
``On the holographic dual of a topological symmetry operator,''
\href{https://journals.aps.org/prd/abstract/10.1103/PhysRevD.110.046007}{Phys. Rev. D \textbf{110} (2024) no.4, 046007,}
\href{https://arxiv.org/abs/2401.09538}{[arXiv:2401.09538 [hep-th]].}
\bibitem{Heckman:2025lmw}
J.~J.~Heckman, M.~H{\"u}bner and C.~Murdia,
``Symmetry Theories, Wigner's Function, Compactification, and Holography,''
\href{https://arxiv.org/abs/2505.23887}{[arXiv:2505.23887 [hep-th]].}
\bibitem{Apruzzi:2025hvs}
F.~Apruzzi, N.~Dondi, I.~Garc{\'\i}a Etxebarria, H.~T.~Lam and S.~Schafer-Nameki,
``Symmetry TFTs for Continuous Spacetime Symmetries,''
\href{https://arxiv.org/abs/2509.07965}{[arXiv:2509.07965 [hep-th]].}
\bibitem{Bonetti:2025dvm}
F.~Bonetti, M.~Del Zotto and R.~Minasian,
``SymTFT for Continuous Symmetries: Non-linear Realizations and Spontaneous Breaking,''
\href{https://arxiv.org/abs/2509.10343}{[arXiv:2509.10343 [hep-th]].}
\bibitem{Baguet:2015sma}
A.~Baguet, O.~Hohm and H.~Samtleben,
``Consistent Type IIB Reductions to Maximal 5D Supergravity,''
\href{https://journals.aps.org/prd/abstract/10.1103/PhysRevD.92.065004}{Phys. Rev. D \textbf{92} (2015) no.6, 065004}
\href{https://arxiv.org/abs/1506.01385}{[arXiv:1506.01385 [hep-th]].}
\bibitem{Witten:1998wy}
E.~Witten,
``AdS/CFT correspondence and topological field theory.,''
\href{https://iopscience.iop.org/article/10.1088/1126-6708/1998/12/012}{JHEP \textbf{12} (1998), 012}
\href{https://arxiv.org/abs/hep-th/9812012}{[arXiv:hep-th/9812012 [hep-th]].}
\bibitem{Bergman:2024aly}
O.~Bergman, E.~Garcia-Valdecasas, F.~Mignosa and D.~Rodriguez-Gomez,
``Non-BPS branes and continuous symmetries,''
\href{https://link.springer.com/article/10.1007/JHEP02(2025)066}{JHEP \textbf{02} (2025), 066}
\href{https://arxiv.org/abs/2407.00773}{[arXiv:2407.00773 [hep-th]].}

\bibitem{Calvo:2025kjh}
H.~Calvo, F.~Mignosa and D.~Rodriguez-Gomez,
``Continuous symmetry defects and brane/anti-brane systems,''
\href{https://link.springer.com/article/10.1007/JHEP06(2025)196}{JHEP \textbf{06} (2025), 196}
\href{https://arxiv.org/abs/2503.04892}{[arXiv:2503.04892 [hep-th]].}

\bibitem{Calvo:2025usj}
H.~Calvo, F.~Mignosa and D.~Rodriguez-Gomez,
``R-symmetries, anomalies and non-invertible defects from non-BPS branes,''
\href{https://arxiv.org/abs/2506.13859}{[arXiv:2506.13859 [hep-th]].}




\end{thebibliography}
\end{document}